\documentclass[10pt,conference]{IEEEtran}
\IEEEoverridecommandlockouts
\usepackage{cite}
\usepackage{amsmath,amssymb,amsfonts}
\usepackage{booktabs}
\usepackage{multirow}
\usepackage{tabularx}
\usepackage{algorithmic}
\usepackage{graphicx}
\usepackage{textcomp}
\usepackage{xcolor}
\usepackage{url}
\usepackage[most]{tcolorbox}
\usepackage{booktabs}
\def\BibTeX{{\rm B\kern-.05em{\sc i\kern-.025em b}\kern-.08em
    T\kern-.1667em\lower.7ex\hbox{E}\kern-.125emX}}
\begin{document}
\bstctlcite{BSTcontrol}

\newtcolorbox{insightbox}{
  colback=gray!8,
  colframe=black,
  boxrule=1.2pt,
  arc=5pt,
  left=6pt,
  right=6pt,
  top=5pt,
  bottom=5pt,
  boxsep=0pt,
  width=\linewidth,
  before skip=6pt,
  after skip=6pt,
  fontupper=\normalsize
}

\title{SequenceFI: Non-intrusive Temporal Fault Injection for Microservice Systems}

\author{
\IEEEauthorblockN{Yuzhen Tan}
\IEEEauthorblockA{
\textit{School of Computer Science}\\
Wuhan University\\
Wuhan, Hubei, China\\
tanyuzhen@whu.edu.cn
}
\and
\IEEEauthorblockN{Jian Wang}
\IEEEauthorblockA{
\textit{School of Computer Science}\\
Wuhan University\\
Zhongguancun Laboratory\\
Wuhan, Hubei, China\\
jianwang@whu.edu.cn
}
\and
\IEEEauthorblockN{Bing Li}
\IEEEauthorblockA{
\textit{School of Computer Science}\\
Wuhan University\\
Zhongguancun Laboratory\\
Wuhan, Hubei, China\\
bingli@whu.edu.cn
}
\and
\IEEEauthorblockN{Shaolin Tan}
\IEEEauthorblockA{
Zhongguancun Laboratory\\
Beijing, China\\
shaolintan@hnu.edu.cn
}
}

\maketitle

\begin{abstract}
Fault injection is widely used to evaluate the resilience of microservice systems, where client requests often span multiple services and execution stages. Existing request-level techniques usually control where and what faults are injected, but not when they are activated within a distributed execution. This limitation makes it difficult to reproduce timing-dependent failures, such as failures after state-changing side effects, order-sensitive concurrent responses, and partial failures among repeated downstream calls. This paper presents SequenceFI, a non-intrusive framework for temporal fault injection in microservice systems. SequenceFI observes message-level send and receive events, propagates compact temporal evidence along request executions, and triggers faults only when occurrence-sensitive temporal guards are satisfied. It further synthesizes temporal guards from traces, reducing the need for exhaustive enumeration of temporal fault-injection configurations, while requiring no modifications to application code or serialization libraries. We implement SequenceFI on Kubernetes and evaluate it on four widely used microservice benchmarks. Across nine temporal-fault scenarios and 450 valid trials, SequenceFI achieves 100.0\% temporal success without premature or multiple injections, finds effective configurations in one attempt on average, and reduces aggregate end-to-end search time by 95.91\% compared with H-Random.
\end{abstract}


\section{Introduction}

Microservice architectures improve scalability, deployment flexibility, and development agility by decomposing applications into independently deployable services. However, a single user request often traverses multiple services through dynamic inter-service interactions, making resilience depend on the behavior of the entire distributed execution. Chaos engineering is widely used to assess such resilience by deliberately introducing adverse conditions and observing whether systems maintain or recover expected behavior~\cite{ChaosEngineeringBasic, Netflix-ICSE, ChaosMonkey, ChaosMesh, ChaosBlade}. Fault injection (FI) provides a key experimental mechanism for chaos engineering and resilience testing by perturbing requests, services, networks, or containers.

Existing request-level FI techniques~\cite{Netflix-ICSE, MicroFI} can specify \emph{where} a fault is injected and \emph{what} fault is applied, such as failing an API request or response. However, many microservice resilience failures also depend on \emph{when} the fault occurs within a distributed execution. For example, dropping a payment response after the payment has been charged creates a different recovery problem from failing the payment request before the charge is issued; a risk-check failure may expose a bug only if it arrives after another concurrent response has already caused the caller to commit success; and a product-page aggregation bug may appear only when a specific subset of homogeneous downstream responses fails. These examples capture three representative temporal constraints: execution phase, relative response order, and occurrence or cardinality among repeated calls. We formalize these constraints in Section~\ref{sec:temporal-fault-scenarios}. Without temporal control, an FI experiment may trigger too early, affect the wrong occurrence, or collapse a partial failure into an all-or-nothing failure, thereby missing the intended resilience defect and creating false confidence in recovery behavior. Testing these scenarios requires temporal fault injection (TFI), where a fault is activated only when specified execution events or occurrence conditions have been satisfied.

As discussed in Section~\ref{sec:motivation-tfi-nonidempotent}, the need for temporal control is not limited to rare corner cases. Under a method-level approximation, our analysis of eight public API specification corpora identifies 19{,}701 of 38{,}631 operations (51.0\%) as potentially state-changing. For these operations, a failure before a downstream side effect and the same failure after the side effect can impose fundamentally different recovery obligations, especially under retries, compensation, and partial-failure handling. More generally, distributed workflows involving concurrent or repeated downstream calls can exhibit order-sensitive and occurrence-sensitive behavior. These observations make temporal position an essential dimension of practical microservice fault injection.

Prior work has explored TFI for microservices. 3MileBeach~\cite{3milebeach}, the most directly related TFI system we are aware of, attaches temporal prerequisites to temporal fault-injection configurations (TFICs). However, it achieves temporal control via serialization-layer instrumentation and explores TFICs via randomized enumeration. As a result, practical cloud-native TFI still faces two challenges. \textbf{(1) Deployment bottleneck of serialization-layer TFI.} Serialization-layer TFI couples the injector to language-specific libraries and generated code; Section~\ref{sec:motivation-for-sidecar-TFI} reports 170 serialization libraries across nine language families. \textbf{(2) TFIC-generation bottleneck.} TFIC generation must jointly search static targets and temporal guards; Section~\ref{sec:motivation-TFIC-generation} shows a three-event guard space reaches 4.04M candidates. Although lineage-driven FI techniques such as LDFI and FastFI reduce search space by leveraging request lineages and static targets~\cite{ldfi, FastFI}, they do not synthesize temporal guards. 

This paper presents \emph{SequenceFI}, a lightweight non-intrusive framework for practical TFI in microservice systems. SequenceFI places the interposition point on the communication path rather than inside application code or serialization libraries. Its TFIProxy sidecars observe message-level send/receive events, propagate compact temporal evidence along request executions, and inject faults only when both the static target and the required temporal context are satisfied. To avoid exhaustive TFIC enumeration, SequenceFI separates static target selection from temporal guard generation: it derives occurrence-aware targets from request lineages and generates compact \texttt{After} guards from traces that distinguish the intended execution context from earlier matching points.

We implement SequenceFI as a Kubernetes-based prototype for HTTP and gRPC microservices and evaluate it on four benchmarks. Across nine temporal-fault scenarios, SequenceFI achieves 100.0\% temporal success across all valid trials without premature or multiple injections. The post-effect experiments expose benchmark-level recovery weaknesses in state-changing workflows, where repeated client actions after an uncertain outcome can lead to duplicate business operations. For temporal guard generation, SequenceFI identifies an effective configuration in one attempt in all cases and reduces aggregate end-to-end search time by 95.91\% compared with H-Random, a trace-guided randomized baseline.

The main contributions of the paper are summarized as follows:

% To the best of our knowledge, SequenceFI is the first non-intrusive, sidecar-based TFI framework for microservice systems without modifying application code and the first microservice TFI approach that automatically generates effective occurrence-aware TFICs from observed traces.

\begin{itemize}
    \item We propose SequenceFI, a sidecar-based TFI framework that observes message-level events, propagates compact temporal evidence, and triggers faults only when occurrence-aware static targets and execution-context-sensitive temporal guards are jointly satisfied. To the best of our knowledge, SequenceFI is the first non-intrusive TFI framework for microservice systems without modifying application code.

    \item We design a novel trace-guided TFIC generation algorithm that separates static target selection from temporal guard synthesis and derives compact \texttt{After} predicates from distinguishing temporal evidence, thereby avoiding exhaustive or randomized enumeration of temporal configurations.

    \item We evaluate SequenceFI on four microservice benchmarks, showing precise temporal triggering, efficient TFIC generation, and practical deployment with low runtime overhead.
\end{itemize}

\section{Background and Motivation}
\subsection{Background}
% \subsubsection{Microservice Request Executions}
% \label{sec:bg:microservice-execution}
% Modern microservice applications implement a user-visible operation as a distributed execution that spans multiple independently deployed services. A checkout operation, for example, may involve a frontend service, a checkout service, an inventory service, and a payment service. These services interact through inter-service messages, such as RPC or HTTP requests and responses.

% \begin{figure}[t]
%   \centering
%   \includegraphics[width=\linewidth]{ArticleImage/2.1.pdf}
%   \caption{Message-level view of a microservice request execution.}
%   \label{fig:microservice-request-execution}
% \end{figure}

% As shown in Fig.~\ref{fig:microservice-request-execution}, we model a microservice execution as a trace of observable message-level events. For each inter-service call, the trace contains four boundary events: request send, request receive, response send, and response receive. These boundary events form the event vocabulary used by our temporal model. This message-level view is useful because it exposes the temporal position of a distributed request without requiring the fault injector to understand each service's internal implementation.

% Because these events are ordered in a trace, an execution can also be described by temporal relations among boundary events. For example, one event may occur before, after, or between other events in the same request execution. These relations provide the basic structure for specifying temporal conditions over microservice executions.

\subsubsection{Microservice Request Executions}
\label{sec:bg:microservice-execution}
\begin{figure}[t]
  \centering
  \includegraphics[width=\linewidth]{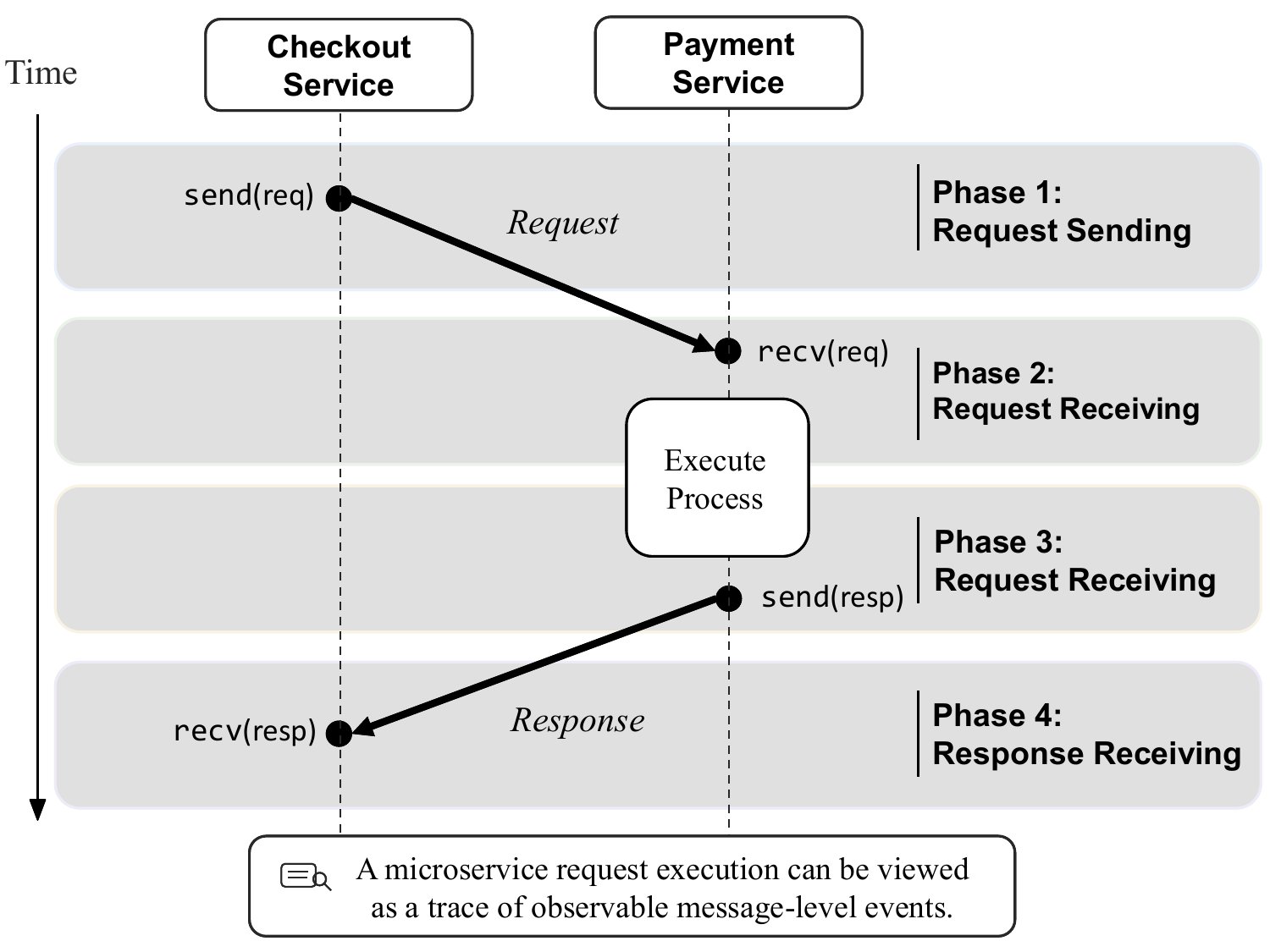}
  \caption{Message-level view of a microservice request execution.}
  \label{fig:microservice-request-execution}
\end{figure}
A user-visible operation in a microservice application may invoke a chain or fan-out of calls across independently deployed services. We represent such an execution as a trace of observable message-level boundary-event occurrences. For each inter-service HTTP or RPC call, the trace records four event types: request send, request receive, response send, and response receive, as illustrated in Fig.~\ref{fig:microservice-request-execution}. These events provide a communication-level view of the distributed execution without requiring access to service-internal logic. For an individual API call, the four boundary events follow a causal order from request send to response receive. Events belonging to concurrent branches, however, may not have a unique global order. Temporal conditions can therefore be expressed in terms of whether particular boundary events have occurred before a selected injection point, rather than assuming a total order over all events in the distributed execution.

\subsubsection{Request-Level and Temporal Fault Injection}
\label{sec:bg:rlfi}
\label{sec:bg:tfi}
Request-level fault injection (RLFI) confines a perturbation to selected request executions by specifying a request selector, an injection point or phase, and a fault type. For example, an RLFI configuration may inject an error into requests to, or responses from, a selected service endpoint. Temporal fault injection (TFI) extends RLFI with a temporal predicate over events observed within the same distributed request execution. This predicate determines whether the fault is eligible to be triggered at its static target; in this paper, we refer to it as a \emph{temporal guard}. Thus, a fault is injected only when both the static target is matched, and the temporal guard is satisfied. 3MileBeach realizes such temporal control by attaching temporal prerequisites to fault-injection configurations~\cite{3milebeach}. Accordingly, a temporal fault-injection configuration specifies not only where and what fault to inject, but also the temporal execution context in which the fault becomes eligible.

% \subsubsection{Lineage-Driven Fault Injection}
% \label{sec:bg:ldfi}
% Lineage-Driven Fault Injection (LDFI) selects fault-injection experiments by reasoning backward from successful executions~\cite{ldfi}. Rather than randomly exploring failures, LDFI uses the causal dependencies of a successful outcome to identify combinations of faults that may invalidate that outcome. This makes LDFI useful for reducing the search space of fault-injection experiments.

% Causal lineage records how an observed outcome depends on prior events, messages, and intermediate results. In a microservice execution, lineage can be viewed as a dependency structure over the messages and service actions that contribute to a successful response. LDFI uses this structure to derive fault-injection experiments from the dependencies that support the outcome.

\subsubsection{State-Changing Operations and Outcome Uncertainty}
\label{sec:bg:state-changing}
Retries are a common recovery mechanism in microservice systems, but their correctness depends on whether an operation can modify durable state or produce externally visible side effects. For the purposes of this paper, we distinguish \emph{safe operations}, whose intended semantics do not request such state changes, from \emph{state-changing operations}, such as creating an order, reserving inventory, charging a payment method, updating a profile, deleting a resource, or publishing an event.

Safety should not be conflated with idempotence. An operation is \emph{idempotent} if multiple identical requests have the same intended effect as a single request. Therefore, an operation may be state-changing while remaining idempotent; \textsc{PUT} and \textsc{DELETE}, for example, are idempotent under HTTP semantics even though they can modify server-side state~\cite{rfc9110}.

For state-changing operations, an important temporal fault window is the \emph{outcome-uncertain window}: the downstream side effect may already have occurred, while the upstream caller has not yet observed its result. A fault in this window can leave the caller unable to determine whether retrying, compensating, or reporting failure is safe, potentially causing duplicate effects, lost updates, or an inconsistent business state.

\subsection{Motivation}
\label{sec:motivation}
\subsubsection{Temporal Faults in State-Changing Operations}
\label{sec:motivation-tfi-nonidempotent}

\begin{table}[t]
\centering
\footnotesize
\setlength{\tabcolsep}{2pt}
\renewcommand{\arraystretch}{1.25}
\caption{Operation-level prevalence of state-changing operations in eight public API corpora.}
\label{tab:motivation-state-changing}
\begin{tabular*}{\columnwidth}{@{\extracolsep{\fill}}lcccc@{}}
\hline
Corpus & Total & Safe & \shortstack[c]{State-\\Changing} & Percent(\%) \\
\hline
Microsoft Graph v1.0~\cite{msgraph-openapi} & 16422 & 8634 & 7788 & 47.4 \\
Azure REST API Specs~\cite{azure-rest-api-specs} & 8083 & 4090 & 3993 & 49.4 \\
Google APIs Discovery~\cite{google-discovery} & 8016 & 3231 & 4785 & 59.7 \\
Cloudflare API~\cite{cloudflare-openapi} & 3051 & 1475 & 1576 & 51.7 \\
GitHub REST API~\cite{github-rest-api-description} & 1186 & 620 & 566 & 47.7 \\
DigitalOcean API v2~\cite{digitalocean-openapi} & 635 & 331 & 304 & 47.9 \\
Stripe API~\cite{stripe-openapi} & 619 & 274 & 345 & 55.7 \\
Jira Cloud REST API v3~\cite{jira-cloud-openapi} & 619 & 275 & 344 & 55.6 \\
\hline
Total & 38631 & 18930 & 19701 & 51.0 \\
\hline
\end{tabular*}
\vspace{-2ex}
\end{table}

Following the HTTP method semantics specified in RFC~9110~\cite{rfc9110}, we classify \textsc{GET}, \textsc{HEAD}, \textsc{OPTIONS}, and \textsc{TRACE} as safe methods. As a method-level approximation, we classify \textsc{POST}, \textsc{PUT}, \textsc{PATCH}, and \textsc{DELETE} as potentially state-changing because they may modify application state or produce externally visible side effects. As summarized in Table~\ref{tab:motivation-state-changing}, our analysis covers 38{,}631 unique operations from eight public API specification corpora. Of these operations, 19{,}701 (51.0\%) use potentially state-changing methods, with the proportion in each corpus ranging from 47.4\% to 59.7\%.

Such operations can create an \emph{outcome-uncertain window} when a downstream side effect has occurred, but a failure prevents the caller from observing its result. In this window, the caller cannot determine whether retrying, compensating, or reporting failure is safe; an incorrect recovery action may duplicate, lose, or corrupt business state. A conventional injector that triggers before the side effect cannot reproduce this state, even when it targets the same API. Testing such recovery behavior, therefore, requires FI to control not only \emph{where} a fault occurs, but also \emph{when} it occurs relative to the state-changing execution.

\begin{insightbox}
\textbf{Insight 1:} Potentially state-changing methods constitute 51.0\% of the operations in eight public API specification corpora. Because failures before and after a side effect impose different recovery obligations, practical FI should distinguish the temporal window in which a fault occurs.
\end{insightbox}

\subsubsection{Limitations of Serialization-Layer TFI}
\label{sec:motivation-for-sidecar-TFI}
\begin{table}[t]
\centering
\footnotesize
\setlength{\tabcolsep}{4pt}
\renewcommand{\arraystretch}{1.25}
\caption{Language-level diversity of serialization/deserialization libraries and their covered data formats.}
\label{tab:motivation2-serialization-diversity}
\begin{tabular*}{\columnwidth}{@{\extracolsep{\fill}}lcp{5.2cm}@{}}
\hline
Language & Libs & Formats \\
\hline
C/C++$^\dagger$ & 55 & JSON, YAML, Protobuf, MessagePack, FlatBuffers, Cap'n Proto, Thrift, binary \\
C\# / .NET & 34 & JSON, XML, YAML, Protobuf, MessagePack, BSON, binary \\
Java & 21 & JSON, XML, YAML, Protobuf, Avro, MessagePack, Hessian, binary \\
Python & 14 & JSON, YAML, MessagePack, Protobuf, CBOR, Avro, Thrift, Pickle \\
JavaScript / Node.js & 12 & JSON, YAML, XML, Protobuf, CBOR, MessagePack, Avro \\
Kotlin & 11 & JSON, Protobuf, CBOR, YAML, HOCON, Properties \\
Go & 10 & JSON, YAML, Protobuf, CBOR, MessagePack, Avro \\
PHP & 8 & JSON, XML, YAML, Protobuf, MessagePack \\
TypeScript & 5 & JSON, Protobuf, CBOR, MessagePack, Avro \\
\hline
Total & 170 & JSON, XML, YAML, Protobuf, MessagePack, CBOR, Avro, Thrift, binary \\
\hline
\end{tabular*}
\\[2pt]
\footnotesize
$^\dagger$ C/C++ combines the C++ and C topic slices reported by GitHub.
\vspace{-2ex}
\end{table}

\begin{figure*}[t]
\centering
\includegraphics[width=\textwidth]{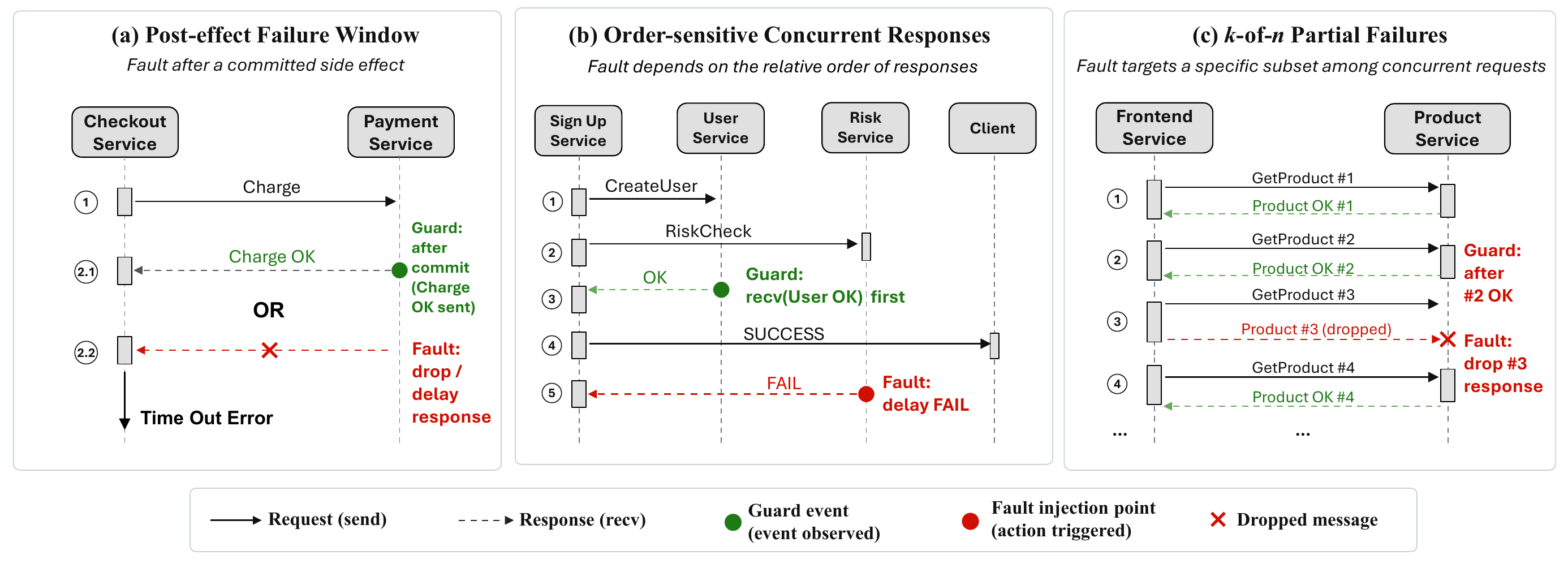}
\caption{Three representative temporal fault patterns in microservices. Event-guarded fault injection uses send/receive events as temporal evidence to target post-effect, order-sensitive, and occurrence-specific failure windows that are difficult to distinguish using static request-level or phase-level fault injection.}
\label{fig:tfi-scenarios}
\vspace{-2ex}
\end{figure*}

The closest existing microservice TFI approach, 3MileBeach, achieves temporal control by instrumenting the message serialization/deserialization layer~\cite{3milebeach}. Although this design avoids modifying application logic, it couples the fault injector to concrete serialization stacks, including language-specific libraries, generated-code frameworks, encoders, and wire formats. In polyglot microservice systems, these stacks can differ across services and evolve independently, making it difficult to identify and maintain uniform instrumentation points.

To characterize this heterogeneity, we use repositories listed under GitHub's serialization-library topic~\cite{github-serialization-topic} as a conservative public indicator of the serialization ecosystem. As shown in Table~\ref{tab:motivation2-serialization-diversity}, the major language slices contain 170 repositories across nine language families and support diverse formats, including JSON, XML, YAML, Protobuf, Avro, Thrift, and custom binary encodings. This evidence does not imply that every library must be individually supported, but it illustrates the recurring adaptation and validation burden of library-level instrumentation as serialization libraries, generated code, runtime behavior, and data formats evolve. These limitations motivate moving the interposition point to a more implementation-independent communication layer.

\begin{insightbox}
\textbf{Insight 2:} Serialization-layer TFI is coupled to heterogeneous and evolving language-specific stacks. A communication-layer proxy can provide temporal interception without modifying application code or serialization libraries.
\end{insightbox}

\subsubsection{Configuration-Space Explosion in TFIC Generation}
\label{sec:motivation-TFIC-generation}
\begin{table}[t]
\centering
\footnotesize
\setlength{\tabcolsep}{4pt}
\renewcommand{\arraystretch}{1.25}
\setlength{\heavyrulewidth}{0.4pt}
\setlength{\lightrulewidth}{0.4pt}
\caption{Naive TFIC enumeration for representative benchmark requests.}
\label{tab:motivation-tfic-enumeration}
\begin{tabular*}{\columnwidth}{@{\extracolsep{\fill}}p{2.4cm}p{2.4cm}cccc@{}}
\toprule
Benchmark & Request & $N_R$ & $r{=}1$ & $r{=}2$ & $r{=}3$ \\
\midrule
Online Boutique
  & \texttt{POST /checkout}
  & 12 & 576 & 13.5K & 207.6K \\
Hotel Reservation
  & \texttt{GET /hotels}
  & 5 & 100 & 950 & 5.7K \\
Sock Shop
  & \texttt{POST /orders}
  & 9 & 324 & 5.7K & 64.3K \\
Train Ticket
  & \texttt{POST /preserve}
  & 25 & 2.5K & 123.8K & 4.04M \\
\bottomrule
\end{tabular*}
\begin{minipage}{0.98\columnwidth}
\footnotesize
% $N_R$: number of APIs invoked by the selected request;
% $r$: number of temporal events in the guard.
K and M denote $10^3$ and $10^6$.
\end{minipage}
\vspace{-2ex}
\end{table}

TFI introduces a configuration burden because a TFIC must specify both a static injection target and a temporal guard. We illustrate this burden using four microservice benchmarks. For each benchmark, we select the request trace containing the largest number of inter-service API invocations, thereby examining the most temporally complex observed request. For a request $R$ invoking $N_R$ APIs, each API exposes up to four boundary events, yielding $4N_R$ candidate temporal events. Assuming an $r$-event guard is a conjunction of $r$ distinct events whose order within the guard is irrelevant, a naive enumerator explores $N_R\binom{4N_R}{r}$ configurations, where $N_R$ accounts for static targets and $\binom{4N_R}{r}$ for temporal guards.

This estimate remains conservative because it ignores fault-type multiplicity, repeated occurrences, request/response phase choices, retries, and concurrent fan-out. Even under this simplified model, Table~\ref{tab:motivation-tfic-enumeration} shows that naive enumeration reaches 207.6K candidates for Online Boutique and 4.04M for Train Ticket with only three temporal events. Thus, practical TFI requires algorithmic generation of static targets and temporal guards rather than relying on randomized TFIC enumeration.

\begin{insightbox}
\textbf{Insight 3:} Even under a simplified distinct-event model, the TFIC search space grows to millions of candidates for a three-event guard. This motivates algorithmic static-target selection and temporal-guard generation.
\end{insightbox}

\section{Temporal Fault Patterns in Microservices}
\label{sec:temporal-fault-scenarios}

% Traditional fault injection can specify where a fault is injected and what fault is applied, but it often cannot constrain the execution context in which the fault becomes active. As a result, failures that depend on committed side effects, response order, or partial failures among homogeneous calls are difficult to reproduce reliably. 
Fig.~\ref{fig:tfi-scenarios} illustrates three temporal fault scenarios in microservice executions. From these scenarios, we identify three representative temporal fault patterns that require event-guarded injection to reproduce timing-dependent failure conditions.

\subsection{Post-Effect Failure Window}
In the checkout-payment scenario in Fig.~\ref{fig:tfi-scenarios}(a), Payment may charge the user, but its response is lost before reaching Checkout. Checkout observes a timeout despite the completed charge, creating an outcome-uncertain state in which an unsafe retry may produce a duplicate payment. This scenario instantiates a post-effect failure pattern, where a failure occurs after a state-changing operation has taken effect but before its caller observes the result. Static request-level FI cannot construct this window: failing Checkout prevents the charge, while failing Payment before completion models an unsuccessful charge rather than a successful charge with a lost result. Static phase matching may be insufficient when repeated calls or intervening events make the occurrence of the response ambiguous. TFI instead conditions activation on message events, allowing Payment to complete before faulting its response.

\subsection{Order-Sensitive Concurrent Responses}
In the signup scenario in Fig.~\ref{fig:tfi-scenarios}(b), UserService may return a successful CreateUser response before RiskService returns a failure. If the caller commits success after CreateUser, the later RiskCheck failure may arrive too late to prevent an externally visible success. This scenario instantiates an order-sensitive concurrent-response pattern, where application behavior depends on the arrival order of concurrent responses. Static FI can make RiskCheck fail, but cannot ensure that the failure occurs after CreateUser has influenced the caller in the intended interleaving. TFI uses ordering evidence to activate the RiskCheck failure after the CreateUser success has been observed, making the relevant interleaving reproducible.

\subsection{$k$-of-$n$ Partial Failures}
In the aggregation scenario in Fig.~\ref{fig:tfi-scenarios}(c), a frontend may issue multiple homogeneous GetProduct requests while rendering a page. Its aggregation logic may behave differently when all calls succeed, all fail, or only selected responses fail. This scenario instantiates a $k$-of-$n$ partial-failure pattern, where correctness depends on exactly which subset of homogeneous or concurrent requests fails. Route-based static FI is too coarse-grained because it may affect every matching GetProduct call rather than a designated occurrence. TFI uses occurrence counts, request identities, or trace-level evidence to target selected calls, enabling precise $k$-of-$n$ failure states.

The three patterns are not intended to form an exhaustive or mutually exclusive taxonomy of microservice failures. Instead, they characterize three primitive temporal constraints needed to reproduce the motivating scenarios: phase constraints, ordering constraints, and cardinality constraints. More complex temporal fault scenarios can be modeled by composing these constraints across multiple events and requests. This observation motivates our method: observing send/receive events, evaluating temporal guards over these events, and triggering faults only when the corresponding guard is satisfied.

\section{Method}
\begin{figure}[t]
  \centering
  \includegraphics[width=\linewidth]{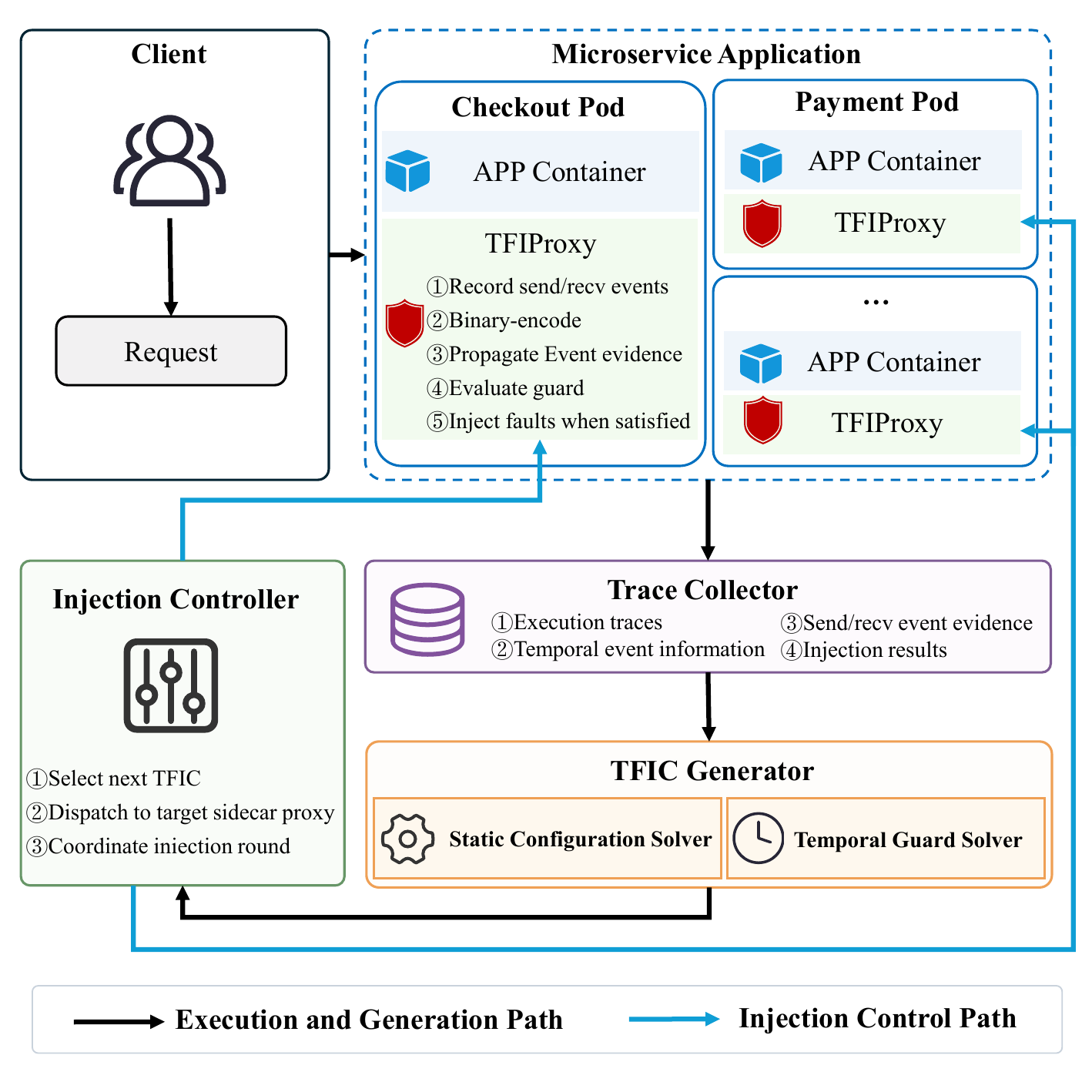}
  \caption{Overview of SequenceFI.}
  \label{fig:sequencefi-overview}
  \vspace{-2ex}
\end{figure}
\subsection{Approach Overview}
Figure~\ref{fig:sequencefi-overview} presents the architecture of SequenceFI, which comprises four components: TFIProxy sidecars deployed alongside microservice pods, a trace collector, a TFIC generator, and an injection controller. A temporal fault-injection configuration (TFIC) consists of two parts: a static configuration and a temporal guard. The static configuration specifies \emph{where} and \emph{what} to inject, including the target API, the occurrence index, the injection phase, and the fault type. The temporal guard specifies \emph{when} the fault should become active by constraining the execution context through previously observed events. A fault is injected only when both the static target and temporal guard are satisfied. SequenceFI observes only message-level send/receive events through TFIProxy sidecars, requiring no modification to application code or serialization libraries. During execution, each TFIProxy records message events, propagates compact binary-encoded temporal evidence across service boundaries, evaluates temporal guards online, and triggers the configured fault only when the required temporal context has been reached.

SequenceFI operates through two interacting paths. Along the \emph{execution and generation path} (black arrows), client requests are executed, traces and injection outcomes are collected, and the TFIC generator derives candidate configurations. Along the \emph{injection control path} (blue arrows), the injection controller distributes the generated TFICs to the corresponding TFIProxy sidecars for the next injection round. Injection results are continuously fed back to the TFIC generator, allowing SequenceFI to iteratively refine configurations until effective temporal fault injections are identified.

\subsection{TFIC Generation}
SequenceFI generates TFICs in a feedback-driven manner. Given a workload, it first executes the request without fault injection and collects the corresponding trace. Each trace is represented as an ordered sequence of observable send/receive events. Based on the collected traces and previous injection results, SequenceFI generates a TFIC in two steps: static configuration solving and temporal guard generation. The former determines the target API, the injection phase, and the fault type, while the latter determines the temporal condition under which the fault should be triggered.

\textbf{Static configuration solving.}
A TFIC is decomposed into a static configuration and a temporal guard. The static configuration is defined as:
\begin{equation}
\sigma=\langle api, k, phase, faultType\rangle,
\label{eq:static-config}
\end{equation}
where $api$ denotes the target service endpoint or RPC method to be fault-injected, $k$ denotes the local occurrence index of the specified $(api, phase)$ operation within a request execution, $phase \in \{req, resp\}$ specifies whether the fault is injected at the request or response side, and $faultType$ denotes the fault action, such as delay, drop, error, or abort. Here, $k$ is counted only among runtime operations that match the same $api$ and $phase$; it does not encode a global temporal order over all events in the trace.

SequenceFI adapts FastFI~\cite{FastFI}, which builds on LDFI, to derive candidate static configurations from request lineages. Each candidate variable represents a possible occurrence-specific static injection point, identified by the target API, the local occurrence index of the corresponding API-phase operation, the injection phase, and the fault type. The solver searches for a minimal set of candidates that can affect the client-level request outcome. 
% This step only reasons about the target API, the local occurrence index, the injection phase, and the fault type, leaving the temporal triggering condition to the next step. This separation avoids enumerating all combinations of static targets and temporal guards upfront.

\textbf{Temporal guard generating.}
SequenceFI restricts guards to a uniform \texttt{After} form over conjunctions of occurrence-count atoms. This form covers three temporal fault patterns in Section~\ref{sec:temporal-fault-scenarios} because each can be expressed as prerequisite evidence before a selected target occurrence: post-effect faults require a success-response event, order-sensitive faults require an observed competing response, and $k$-of-$n$ faults require preceding homogeneous occurrences. The static configuration selects the target API, phase, and occurrence, while the \texttt{After} guard checks event-count evidence. This monotonic design enables compact evidence propagation and distinguishes the target occurrence from premature trigger points.

For each static configuration $\sigma$ derived for a client-level request $r$, SequenceFI generates an \texttt{After} guard that enables the fault after the required temporal evidence has been observed. Let $\tau_{\text{new}}$ denote the trace from the latest execution of request $r$, and let $\mathcal{T}_r$ denote the set of observed traces for the same request, including $\tau_{\text{new}}$. We model each trace $\tau\in\mathcal{T}_r$ as a time-ordered sequence of observed events, $\tau=\langle e_1,\ldots,e_m\rangle$. For an event $e$, its prefix count at position $i$ is:
\begin{equation}
\text{cnt}_{\tau}(e,i)=|\{j<i \mid e_j=e\}|.
\label{eq:prefix-count}
\end{equation}

An atomic guard condition is written as $(e,t)$, meaning that event $e$ has occurred at least $t$ times when the guard is evaluated at the current injection point. Let $i_\sigma(\tau)$ denote the position in trace $\tau$ of the $k$-th runtime occurrence that matches the $api$ and $phase$ specified by $\sigma$. SequenceFI constructs candidate atoms that are true at the target occurrence across all traces in $\mathcal{T}_r$:
\begin{equation}
\mathcal{U}_{r,\sigma}=
\{(e,t_e) \mid t_e=\min_{\tau\in\mathcal{T}_r}
\text{cnt}_{\tau}(e,i_\sigma(\tau)), \ t_e>0\}.
\label{eq:candidate-atoms}
\end{equation}

Since $\sigma$ includes the local occurrence index $k$, it identifies the target position $i_\sigma(\tau)$ in each trace $\tau$, i.e., the $k$-th occurrence of the specified $(api, phase)$ operation. However, the same $(api, phase)$ operation may occur before $i_\sigma(\tau)$; we denote each earlier occurrence position by $p$, with $p<i_\sigma(\tau)$. These earlier occurrences are potential premature trigger points: if a guard does not exclude them, the fault may be enabled before the execution reaches the temporal context. To prevent such triggering, SequenceFI compares the target position with each earlier occurrence position $p$ in each trace $\tau\in\mathcal{T}_r$ and computes a distinguishing set:
\begin{equation}
D_{\tau}(p,i_\sigma(\tau))=
\{(e,t)\in\mathcal{U}_{r,\sigma}\mid \text{cnt}_{\tau}(e,p)<t\}.
\label{eq:distinguishing-set}
\end{equation}
Each atom in $D_{\tau}(p,i_\sigma(\tau))$ represents temporal evidence that is satisfied at the target position $i_\sigma(\tau)$ but not at the earlier occurrence $p$. Therefore, choosing at least one atom from this set ensures that the guard remains false at $p$. By selecting atoms that distinguish every earlier occurrence of the same $(api, phase)$ operation from the target position across the observed traces, SequenceFI constructs an \texttt{After} guard that prevents premature triggering and enables the fault only after the required temporal evidence has been observed.

SequenceFI then reduces guard generation to a minimum hitting set problem:
\begin{equation}
G^*=\arg\min_{G\subseteq\mathcal{U}} |G|
\quad
\text{s.t.}\quad
\forall D\in\mathcal{F}, G\cap D\neq\emptyset ,
\label{eq:hitting-set}
\end{equation}
where $\mathcal{F}$ is the collection of all distinguishing sets across traces. The resulting temporal guard is:
\begin{equation}
g=\texttt{After}\left(\bigwedge_{a\in G^*}a\right).
\label{eq:temporal-guard}
\end{equation}
% If multiple minimum solutions exist, SequenceFI applies a deterministic tie-breaking rule, such as lexicographic order of event names, to produce a stable guard. If any distinguishing set is empty, the current event vocabulary cannot distinguish the target occurrence from an earlier conflict, and SequenceFI marks the candidate as ambiguous or waits for additional traces.

\textbf{Feedback-driven refinement.}
After a TFIC $\langle \sigma,g\rangle$ is generated, the injection controller dispatches it to the target sidecar proxy for the next injection round. The proxy injects the specified fault only when the request reaches the static configuration $\sigma$ in Equation~\ref{eq:static-config}, and the guard $g$ in Equation~\ref{eq:temporal-guard} is satisfied. SequenceFI then observes the execution result. If the fault exposes a failure, the TFIC is retained as an effective configuration. Otherwise, the new trace and injection result are fed back to the generator, which refines either the static configuration or the temporal guard. This process repeats until no new effective TFICs can be derived under the current workload and exploration budget.

\section{Implementation}
We implemented SequenceFI as a Kubernetes-based prototype that realizes non-intrusive temporal fault injection through a lightweight sidecar runtime, \textit{TFIProxy}. TFIProxy interposes on application-layer inter-service communication and performs temporal fault injection without modifying application code or language-specific libraries. A Kubernetes mutating admission webhook~\cite{Webhook} automatically injects an init container for traffic redirection and the TFIProxy sidecar for request and response interception. TFIProxy supports HTTP/1.x and gRPC over HTTP/2 and transparently forwards traffic that does not match any deployed TFIC to preserve normal service execution.

\begin{figure}[t]
  \centering
  \includegraphics[width=\columnwidth]{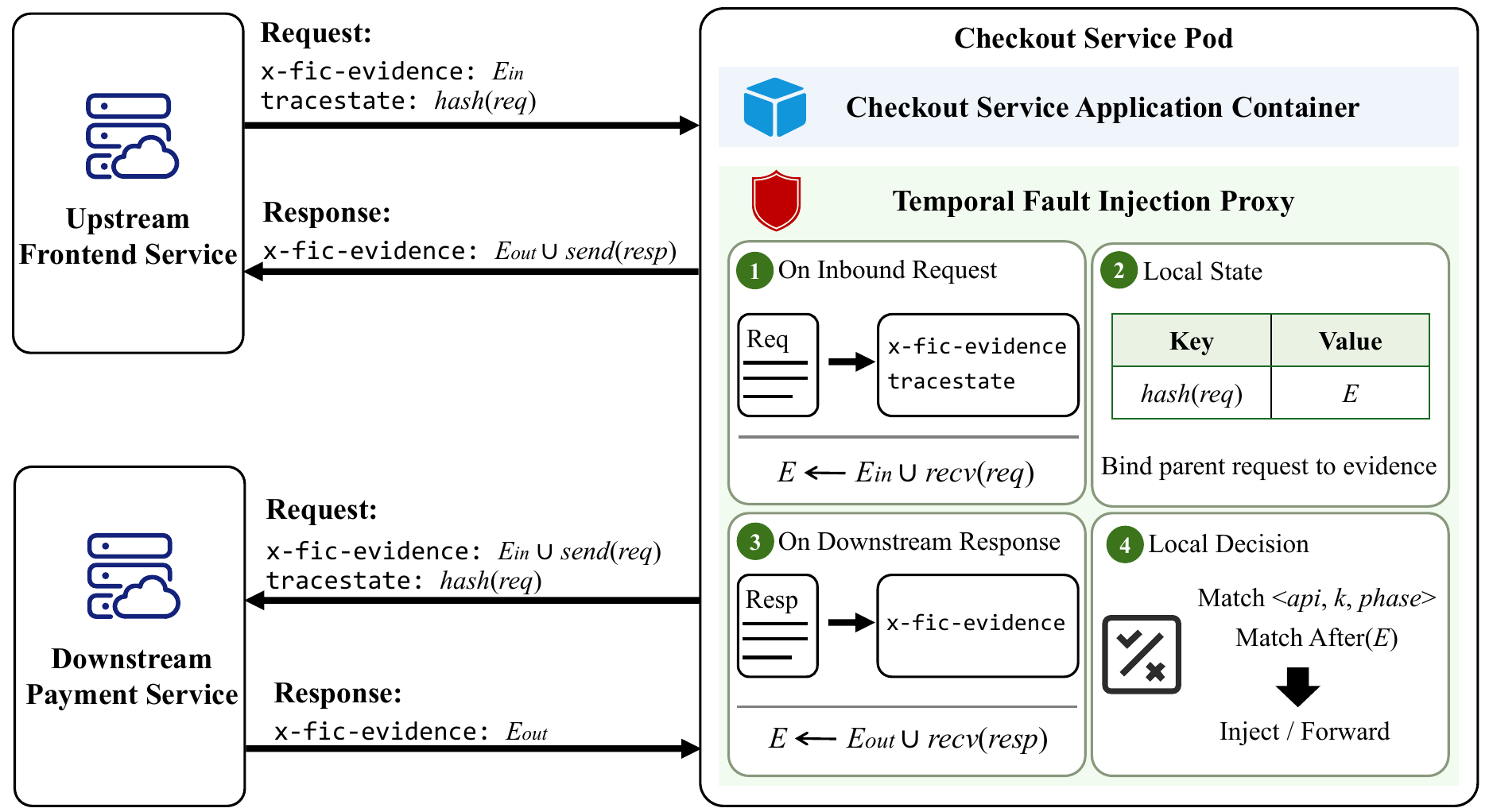}
  \caption{Implementation overview of SequenceFI's TFIProxy sidecar.}
  \label{fig:tfiproxy-evidence-propagation}
  \vspace{-1.2ex}
\end{figure}

As shown in Fig.~\ref{fig:tfiproxy-evidence-propagation}, TFIProxy maintains rule-based runtime configurations derived from TFICs. Each rule specifies the static target, injection phase, fault type, and temporal guard. During execution, TFIProxy observes request/response boundary events and encodes only FIC-relevant events into a compact binary evidence vector. The vector is propagated through the custom \texttt{x-fic-evidence} header in both request and response directions. To preserve causal relationships across service boundaries, W3C Trace Context is used solely for call association: TFIProxy stores a lightweight token in the \texttt{tracestate} field to associate each outbound call with its corresponding inbound parent context~\cite{OpenTelemetry, W3C-context}. Each proxy maintains a local token-to-evidence map, allowing evidence returned from downstream services to be merged into the correct parent request context while avoiding propagation of complete execution histories.

At each interception point, TFIProxy first checks whether the current message matches the static target of the deployed TFIC. If so, it evaluates the temporal guard over the current evidence vector. The configured fault is triggered only when both the static target and temporal guard are satisfied; otherwise, the message is forwarded normally. Our prototype supports delay, drop, connection-reset, and abort-style faults. For aborts, TFIProxy returns HTTP 503 for HTTP traffic and gRPC \texttt{UNAVAILABLE} for gRPC calls. TFIProxy records execution traces, event evidence, and injection outcomes, which are fed back to the TFIC generator for subsequent injection rounds.

\section{Evaluation}
We evaluate SequenceFI through the following research questions, covering temporal triggering correctness, TFIC generation efficiency, and runtime deployment overhead.

\begin{itemize}
    \item \textbf{RQ1:} How effectively can SequenceFI trigger faults at the intended temporal windows of representative temporal fault scenarios while avoiding premature triggering?
    \item \textbf{RQ2:} How efficiently can SequenceFI generate effective TFICs compared with trace-guided and unguided randomized approaches?
    \item \textbf{RQ3:} What runtime overhead does SequenceFI introduce in terms of throughput and sidecar resource consumption?
\end{itemize}

\textbf{RQ1} evaluates whether SequenceFI can correctly trigger faults at the intended temporal windows of representative temporal fault scenarios without premature triggering. \textbf{RQ2} evaluates the efficiency of the proposed TFIC generation algorithm by comparing it with a trace-guided randomized baseline and a 3MileBeach-style unguided random enumeration baseline. \textbf{RQ3} evaluates the runtime overhead of the sidecar-based implementation in terms of throughput and sidecar CPU and memory usage.

\subsection{Evaluation Design}
\subsubsection{Benchmarks}
We use four widely used microservice benchmarks: Online Boutique~\cite{OnlineBoutique}, Hotel Reservation~\cite{hotel}, Sock Shop~\cite{SockShop}, and Train Ticket~\cite{TrainTicket}. These benchmarks are well-suited for evaluating temporal fault injection because client-level requests typically span multiple services, comprise multiple execution stages, and often involve concurrent or repeated downstream calls, thereby providing representative execution contexts for temporal fault scenarios in this paper.

For each benchmark, we select representative business requests whose executions can instantiate the temporal fault patterns discussed in Section~\ref{sec:temporal-fault-scenarios}. We use these requests to evaluate whether our approach can trigger the intended temporal fault scenarios without premature activation, to collect traces for temporal guard generation, and to measure the runtime overhead of sidecar-based fault injection under realistic workloads.

\subsubsection{Evaluation Environment}
All experiments were conducted on a six-node Kubernetes v1.29.15 cluster comprising one control-plane node (Intel Xeon E3-1240 v5, 16 GB RAM) and five worker nodes (Intel Core i7-4790, 16 GB RAM each), all running Ubuntu 22.04. Evaluation scripts used Python 3.9.

\begin{table*}[!t]
\centering
\caption{Temporal-window triggering by representative pattern. Values are percentages over valid trials.}
\label{tab:rq1-scenario}
\scriptsize
\setlength{\tabcolsep}{1.7pt}
\renewcommand{\arraystretch}{1.06}
\begin{tabular*}{\textwidth}{@{\extracolsep{\fill}}p{0.15\textwidth}cc
c
cccccc
cccccc@{}}
\toprule
\multirow{2}{0.15\textwidth}{\raisebox{-2ex}{\begin{tabular}[c]{@{}c@{}}Representative\\Pattern\end{tabular}}} &
\multirow{2}{4.5em}{\raisebox{-0.8ex}{\#Scenarios}} &
\multirow{2}{3.5em}{\raisebox{-0.8ex}{\#Trials}} &
\multicolumn{1}{c}{SequenceFI} &
\multicolumn{6}{c}{Static-Req} &
\multicolumn{6}{c}{Static-Phase} \\
\cmidrule(lr){4-4}\cmidrule(lr){5-10}\cmidrule(lr){11-16}
& & & \emph{TS} & \emph{TS} & \emph{CW} & \emph{PS} & \emph{Prem} & \emph{Miss} & \emph{Mult} & \emph{TS} & \emph{CW} & \emph{PS} & \emph{Prem} & \emph{Miss} & \emph{Mult} \\
\midrule
Post-effect & 3 & 150 & \textbf{100.0} & 0.0 & 0.0 & 0.0 & 100.0 & 100.0 & 0.0 & 100.0 & 100.0 & 100.0 & 0.0 & 0.0 & 0.0 \\
Order-sensitive & 3 & 150 & \textbf{100.0} & 0.0 & 0.0 & 0.0 & 100.0 & 100.0 & 0.0 & 66.7 & 66.7 & 66.7 & 33.3 & 20.0 & 0.0 \\
$k$-of-$n$ & 3 & 150 & \textbf{100.0} & 0.0 & 0.0 & 0.0 & 100.0 & 100.0 & 33.3 & 0.0 & 32.7 & 28.7 & 71.3 & 34.0 & 33.3 \\
\midrule
Overall & 9 & 450 & \textbf{100.0} & 0.0 & 0.0 & 0.0 & 100.0 & 100.0 & 11.1 & 55.6 & 66.4 & 65.1 & 34.9 & 18.0 & 11.1 \\
\bottomrule
\end{tabular*}
\vspace{-1.2ex}
\end{table*}

\begin{table*}[!t]
\centering
\caption{Temporal-window triggering for individual scenarios. Values are percentages over valid trials.}
\label{tab:rq1-specify}
\scriptsize
\setlength{\tabcolsep}{0.55pt}
\renewcommand{\arraystretch}{1.02}
\begin{tabular*}{\textwidth}{@{\extracolsep{\fill}}
>{\tiny\raggedright\arraybackslash}p{0.085\textwidth}
>{\tiny\raggedright\arraybackslash}p{0.155\textwidth}
>{\tiny\raggedright\arraybackslash}p{0.075\textwidth}
@{\hspace{2.5pt}}c
c
c
cccccc
cccccc@{}}
\toprule
\multirow{2}{0.085\textwidth}{\raisebox{-0.8ex}{\scriptsize Benchmark}} &
\multirow{2}{0.085\textwidth}{\raisebox{-0.8ex}{\scriptsize Scenario}} &
\multirow{2}{0.075\textwidth}{\raisebox{-3.6ex}{\scriptsize \begin{tabular}[c]{@{}c@{}}Temporal\\Pattern\end{tabular}}} &
\multirow{2}{1.2em}{\raisebox{-0.8ex}{\scriptsize $k$}} &
\multirow{2}{2.2em}{\raisebox{-0.8ex}{\scriptsize \#Trials}} &
\multicolumn{1}{c}{SequenceFI} &
\multicolumn{6}{c}{Static-Req} &
\multicolumn{6}{c}{Static-Phase} \\
\cmidrule(lr){6-6}\cmidrule(lr){7-12}\cmidrule(lr){13-18}
& & & & & \emph{TS} & \emph{TS} & \emph{CW} & \emph{PS} & \emph{Prem} & \emph{Miss} & \emph{Mult} & \emph{TS} & \emph{CW} & \emph{PS} & \emph{Prem} & \emph{Miss} & \emph{Mult} \\
\midrule
Online Boutique & Payment post-effect & Post-effect & 1 & 50 & \textbf{100.0} & 0.0 & 0.0 & 0.0 & 100.0 & 100.0 & 0.0 & 100.0 & 100.0 & 100.0 & 0.0 & 0.0 & 0.0 \\
Online Boutique & Payment after shipping quote & Order-sensitive & 1 & 50 & \textbf{100.0} & 0.0 & 0.0 & 0.0 & 100.0 & 100.0 & 0.0 & 100.0 & 100.0 & 100.0 & 0.0 & 0.0 & 0.0 \\
Online Boutique & Cart GetProduct response occurrence & $k$-of-$n$ & 2 & 50 & \textbf{100.0} & 0.0 & 0.0 & 0.0 & 100.0 & 100.0 & 0.0 & 0.0 & 0.0 & 0.0 & 100.0 & 34.0 & 0.0 \\
Hotel Reservation & Reservation after user check & Order-sensitive & 1 & 50 & \textbf{100.0} & 0.0 & 0.0 & 0.0 & 100.0 & 100.0 & 0.0 & 100.0 & 100.0 & 100.0 & 0.0 & 0.0 & 0.0 \\
Sock Shop & Payment post-effect & Post-effect & 1 & 50 & \textbf{100.0} & 0.0 & 0.0 & 0.0 & 100.0 & 100.0 & 0.0 & 100.0 & 100.0 & 100.0 & 0.0 & 0.0 & 0.0 \\
Sock Shop & User subresource response & $k$-of-$n$ & 2 & 50 & \textbf{100.0} & 0.0 & 0.0 & 0.0 & 100.0 & 100.0 & 100.0 & 0.0 & 98.0 & 86.0 & 14.0 & 2.0 & 100.0 \\
Train Ticket & Trip-search post-effect & Post-effect & 1 & 50 & \textbf{100.0} & 0.0 & 0.0 & 0.0 & 100.0 & 100.0 & 0.0 & 100.0 & 100.0 & 100.0 & 0.0 & 0.0 & 0.0 \\
Train Ticket & Route after first route response & Order-sensitive & 2 & 50 & \textbf{100.0} & 0.0 & 0.0 & 0.0 & 100.0 & 100.0 & 0.0 & 0.0 & 0.0 & 0.0 & 100.0 & 60.0 & 0.0 \\
Train Ticket & Route response occurrence & $k$-of-$n$ & 3 & 50 & \textbf{100.0} & 0.0 & 0.0 & 0.0 & 100.0 & 100.0 & 0.0 & 0.0 & 0.0 & 0.0 & 100.0 & 66.0 & 0.0 \\
\bottomrule
\end{tabular*}
\vspace{-1.2ex}
\end{table*}

\subsection{RQ1: Effectiveness of Temporal Fault Triggering}
\textbf{Experimental Setup.} RQ1 evaluates whether SequenceFI can trigger representative temporal fault patterns at their intended execution windows while avoiding premature triggering at earlier temporal positions or homogeneous occurrences. We study nine scenarios across four benchmarks, covering post-effect, order-sensitive, and $k$-of-$n$ homogeneous failures. For each scenario, we first collect 10 fault-free traces for temporal guard generation and then perform 50 independent injection trials per method. We compare SequenceFI with two static baselines: \textbf{Static-Req}, which matches only the target API, and \textbf{Static-Phase}, which additionally distinguishes request and response phases but does not support temporal guards. Overall, each method contributes 450 valid trials.

\textbf{Metrics.} Each metric is defined as a Boolean condition for every valid trial and reported as the percentage of valid trials satisfying that condition. The primary metric is temporal success (\emph{TS}), which is satisfied if and only if a fault is triggered exactly once within the intended temporal window of the target scenario: 
$\mathit{TS} =
\mathit{CW}\land
\mathit{PS}\land
\neg\mathit{Prem}\land
\neg\mathit{Miss}\land
\neg\mathit{Mult},$ 
where correct-window (\emph{CW}) is satisfied if at least one injection reaches the intended API, phase, and occurrence; pattern-satisfaction (\emph{PS}) is satisfied if the injected fault satisfies the pattern-specific temporal semantics; premature (\emph{Prem}) is true if at least one injection occurs before the intended temporal window; missed (\emph{Miss}) is true if no injection occurs within the intended temporal window when that window remains observable or can be aligned with a clean execution; and multiple (\emph{Mult}) is true if more than one injection occurs within the same client request. These metrics are not mutually exclusive and are reported to explain why temporal triggering succeeds or fails.
For the post-effect pattern, \emph{PS} is satisfied only if the downstream state change occurs before the caller observes the injected failure. For the order-sensitive pattern, \emph{PS} is satisfied only if the required ordering evidence has been established before fault activation. For the $k$-of-$n$ pattern, \emph{PS} is satisfied only if the fault is injected at the intended homogeneous occurrence without activation at any earlier occurrence. Unlike \emph{TS}, \emph{PS} does not require uniqueness; therefore, additional injections after the intended occurrence preserve \emph{PS} but are reflected by \emph{Mult}.

\textbf{Results.}
Tables~\ref{tab:rq1-scenario} and~\ref{tab:rq1-specify} report aggregate and specific instance results. SequenceFI achieves 100.0\% \emph{TS} across all trials and does not produce premature or multiple injections. This shows that the synthesized temporal guards enable faults only at intended temporal windows. Static-Req fails to trigger any intended temporal window. Its injections occur at request-side events, before post-effect, response-ordering, or occurrence-specific evidence can be held. As a result, Static-Req obtains 0.0\% \emph{TS} for three representative patterns, and Static-Req trials are premature and missed with respect to the intended reference window. This confirms that request-only static matching cannot express the temporal windows.

Static-Phase is more capable but still insufficient without event-level temporal guards. It succeeds in all post-effect trials and in two order-sensitive scenarios where the configured response phase already coincides with the required temporal evidence. However, this success occurs only when the temporal condition degenerates to a static phase match. When ordering evidence or occurrence-level discrimination is required, Static-Phase becomes unreliable: it succeeds in only 66.7\% of order-sensitive trials and 0.0\% of $k$-of-$n$ trials. In Sock Shop, it reaches the user subresource response but injects into multiple homogeneous responses in every trial; in Train Ticket, it fires at earlier route responses and misses the intended occurrence. Overall, Static-Phase improves over Static-Req but still achieves only 55.6\% aggregate \emph{TS},  because phase information alone cannot distinguish execution contexts that require temporal evidence or occurrence-level discrimination.

\begin{insightbox}
  \textbf{Answer to RQ1.} SequenceFI achieves 100.0\% temporal triggering across all evaluated scenarios by combining occurrence-aware static targets with synthesized temporal guards. In contrast, static baselines fail when triggering depends on temporal evidence or occurrence discrimination, confirming that SequenceFI precisely targets post-effect, ordering-dependent, and occurrence-specific windows.
\end{insightbox}

\subsection{RQ2: Performance of TFIC Generation}
\textbf{Experimental Setup.} RQ2 evaluates whether SequenceFI can synthesize effective temporal guards with fewer fault-injection attempts than randomized TFIC generation. An \emph{effective TFIC} triggers the fault at the intended API occurrence and temporal window, without premature, missed, or multiple injections under the RQ1 oracle. Because directly comparable microservice TFI systems are limited, RQ2 focuses on randomized TFIC generation, including the closest prior strategy and a stronger trace-guided variant. We compare SequenceFI with \textbf{3MileBeach-Random}, an unguided enumeration baseline following 3MileBeach~\cite{3milebeach}, and \textbf{H-Random}, which uses the same clean traces as SequenceFI but samples guard atoms heuristically instead of solving the hitting-set problem. All methods share the same benchmark request, static target, phase, fault type, deployment procedure, workload, and oracle.

\textbf{Metrics.}
\emph{Fault Injection Attempts} counts the number of fault-injection attempts required to identify the first effective TFIC. \emph{Solving Time} measures the cumulative local time spent generating candidate TFICs before success. \emph{End-to-end Time} measures the wall-clock search time until the first effective TFIC is confirmed, including guard generation, deployment, execution, trace collection, and oracle checking.

\textbf{Results.} Table~\ref{tab:rq2_benchmark_results} shows that SequenceFI finds an effective TFIC in one attempt for every benchmark, while H-Random and 3MileBeach-Random require 23 and 726 attempts on average, respectively. H-Random reduces exploration with trace-derived heuristics, but still requires many executions in occurrence-sensitive cases such as Sock Shop and Train Ticket. In contrast, SequenceFI distinguishes the intended occurrence from premature trigger points before deployment.
\begin{table}[t]
\centering
\caption{Comparison of TFIC generation efficiency among SequenceFI, H-Random, and 3MileBeach-Random.}
\label{tab:rq2_benchmark_results}
\scriptsize
\setlength{\tabcolsep}{2.2pt}
\renewcommand{\arraystretch}{1.04}
\begin{tabularx}{\columnwidth}{@{}>{\raggedright\arraybackslash}X l c c c@{}}
\toprule
\begin{tabular}[c]{@{}l@{}}Benchmark /\\ \emph{Request}\end{tabular}
& Method
& \begin{tabular}[c]{@{}c@{}}Fault Injection\\Attempts\end{tabular}
& \begin{tabular}[c]{@{}c@{}}Solving\\Time (ms)\end{tabular}
& \begin{tabular}[c]{@{}c@{}}End-to-end\\Time (s)\end{tabular} \\
\midrule
\multirow{3}{*}{\begin{tabular}[t]{@{}l@{}}Online Boutique /\\\emph{Cart workflow}\end{tabular}}
& SequenceFI & \textbf{1} & 18.40 & \textbf{17.27} \\
& H-Rand     & 3 & 61.19 & 56.23 \\
& 3MB-Rand   & 14 & \textbf{5.58} & 253.08 \\
\addlinespace[0.75ex]
\multirow{3}{*}{\begin{tabular}[t]{@{}l@{}}Hotel Reservation /\\\emph{Repeated recommendations}\end{tabular}}
& SequenceFI & \textbf{1} & 2.64 & \textbf{16.92} \\
& H-Rand     & 2 & 13.07 & 38.68 \\
& 3MB-Rand   & 3 & \textbf{0.63} & 57.01 \\
\addlinespace[0.75ex]
\multirow{3}{*}{\begin{tabular}[t]{@{}l@{}}Sock Shop /\\\emph{Order flow}\end{tabular}}
& SequenceFI & \textbf{1} & \textbf{49.70} & \textbf{17.02} \\
& H-Rand     & 39 & 8114.92 & 782.75 \\
& 3MB-Rand   & 1078 & 283.47 & 18242.15 \\
\addlinespace[0.75ex]
\multirow{3}{*}{\begin{tabular}[t]{@{}l@{}}Train Ticket /\\\emph{Left-ticket request}\end{tabular}}
& SequenceFI & \textbf{1} & \textbf{1.61} & \textbf{17.52} \\
& H-Rand     & 48 & 1594.85 & 804.67 \\
& 3MB-Rand   & $1807$ & $441.88$ & $34873.18$ \\
\bottomrule
\end{tabularx}
\vspace{-2ex}
\end{table}

This attempt reduction directly improves end-to-end efficiency. SequenceFI takes 17.18s on average, compared with 420.58s for H-Random and 13{,}356.36s for 3MileBeach-Random. In aggregate across benchmarks, SequenceFI reduces end-to-end temporal-guard search time by 95.91\% over H-Random and 99.86\% over 3MileBeach-Random. These results show that deterministic guard generation outperforms both trace-guided and unguided random enumeration.

\begin{figure*}[!t]
  \centering
  \includegraphics[width=\textwidth]{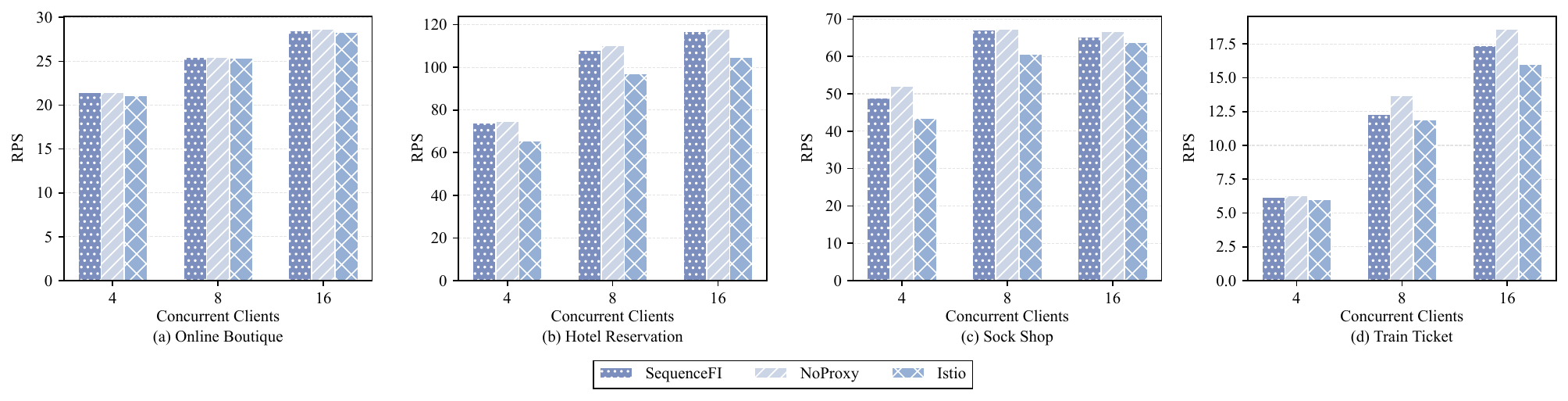}
  \caption{Throughput under different deployment modes.}
  \label{fig:rq3-rps}
  \vspace{-1.5ex}
\end{figure*}

\begin{figure*}[!t]
  \centering
  \includegraphics[width=\textwidth]{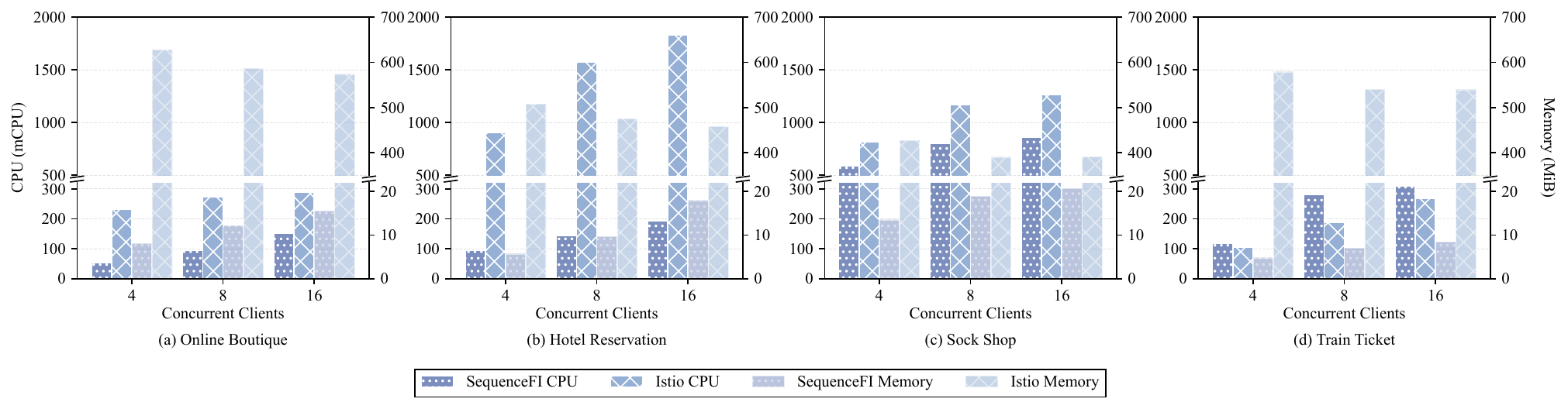}
  \caption{Sidecar CPU and memory usage.}
  \label{fig:rq3-resource}
  \vspace{-1.5ex}
\end{figure*}

\begin{insightbox}
  \textbf{Answer to RQ2.} SequenceFI identifies effective TFICs in a single fault injection attempt across all evaluated benchmarks. Compared with H-Random and 3MileBeach-Random, it reduces the average number of fault injection attempts from 23 and 726 to 1, respectively, while shortening aggregate end-to-end search time by 95.91\% and 99.86\%.
\end{insightbox}

\subsection{RQ3: Overhead of Sidecar-Based TFI}
\textbf{Experimental Setup.} RQ3 evaluates the runtime overhead of SequenceFI when temporal events are observed and guards are evaluated, but no faults are injected. Since SequenceFI adopts a sidecar architecture, we compare it with a widely adopted production-grade sidecar. We compare three deployment modes for throughput: \textbf{NoProxy}, which runs benchmarks without any sidecar; \textbf{SequenceFI}, which deploys the TFIProxy sidecar; and \textbf{Istio}, which deploys an Envoy sidecar. For resource overhead, we compare only the sidecar containers of SequenceFI and Istio because NoProxy has no sidecar. We conduct experiments on four benchmarks with 4, 8, and 16 concurrent clients, yielding 12 workload configurations. Each configuration is repeated at least six times, resulting in 216 measurement trials without request failures.

\textbf{Metrics.} Throughput is measured in requests per second (RPS). Resource overhead is measured by average sidecar CPU and memory consumption in mCPU and MiB, respectively. All resource measurements are container-specific: SequenceFI reports only the TFIProxy sidecar, and Istio reports only the Envoy sidecar, excluding application containers.

\textbf{Throughput overhead.} Figure~\ref{fig:rq3-rps} shows that SequenceFI consistently preserves throughput close to the NoProxy deployment while outperforming the Istio sidecar across all benchmarks and levels of concurrent clients. NoProxy achieves the highest RPS in all 12 workload cases, while SequenceFI consistently ranks between NoProxy and Istio. On average, SequenceFI retains 97.3\% of NoProxy throughput, corresponding to a 2.7\% throughput loss. In comparison, Istio retains 91.5\% of NoProxy throughput, corresponding to an 8.5\% loss. SequenceFI also outperforms Istio in every workload case, improving RPS by 6.6\% on average and up to 12.9\%. 
% The overhead is smallest for Online Boutique, where SequenceFI remains within 1\% of NoProxy across all concurrency levels. The largest SequenceFI throughput loss occurs in Train Ticket, where SequenceFI still retains at least 89.6\% of NoProxy throughput.

\textbf{Sidecar resource overhead.} Figure~\ref{fig:rq3-resource} compares the sidecar CPU and memory usage of SequenceFI and Istio. SequenceFI uses substantially fewer sidecar resources than Istio across most workload cases. Averaged over the 12 cases, SequenceFI consumes 308.20 mCPU, while Istio consumes 743.39 mCPU. Thus, SequenceFI uses only 41.5\% of the CPU used by the Istio sidecar on average. The memory difference is even larger: SequenceFI uses 11.87 MiB on average, while Istio uses 507.95 MiB. Thus, SequenceFI uses only 2.3\% of the memory used by the Istio sidecar on average. This gap reflects the narrower scope of TFIProxy, which performs temporal event tracking and guard evaluation rather than providing a full service-mesh data plane.

\begin{insightbox}
  \textbf{Answer to RQ3.} SequenceFI introduces low runtime overhead. It preserves 97.3\% of NoProxy throughput on average, consistently outperforms the Istio sidecar baseline, and uses substantially less sidecar CPU and memory. These results show that SequenceFI can provide temporal fault-triggering capability with modest throughput cost and lightweight sidecar resource usage.
\end{insightbox}

\section{Discussion}
\subsection{Limitations}
SequenceFI uses OpenTelemetry-compatible W3C Trace Context for cross-service call association and a custom \texttt{x-fic-evidence} header for temporal evidence during request execution~\cite{OpenTelemetry, W3C-context}. Deployments without compatible context propagation require an equivalent call-association mechanism. The current prototype supports HTTP/1.x and gRPC over HTTP/2. Extending SequenceFI to additional communication protocols requires protocol-specific interception, event extraction, and metadata propagation.

SequenceFI currently supports only \texttt{After} guards over positive occurrence-count evidence. This design targets the representative temporal fault patterns studied in this paper, all of which are governed by monotonic enabling conditions: once the required temporal evidence has been observed, the fault remains eligible for injection. Richer operators, such as \texttt{Before} and \texttt{Until}, and finer-grained predicates, such as bounded intervals, event absence, and deadline conditions, could support additional temporal fault scenarios but would require reasoning about non-monotonic temporal conditions. Future work will extend SequenceFI with richer temporal operators, event types, and temporal predicates.

\subsection{Threats to Validity}
\subsubsection{Internal Threats to Validity}
Internal threats concern baseline fairness and implementation artifacts. Because no prior system simultaneously supports sidecar-based deployment and temporal triggering, we use capability-focused baselines: Static-Req and Static-Phase for temporal triggering, and 3MileBeach-Random and H-Random for TFIC generation. We mitigate baseline-related threats by using the same requests, workloads, targets, phases, fault types, deployments, trial settings, and oracles across compared methods. We record traces, propagated evidence, and injection outcomes to validate trials.

\subsubsection{External Threats to Validity}
External threats concern whether our findings generalize beyond the evaluated systems and scenarios. We mitigate this threat by evaluating four microservice benchmarks, Online Boutique, Hotel Reservation, Sock Shop, and Train Ticket~\cite{OnlineBoutique, hotel, SockShop, TrainTicket}, which cover diverse service topologies, request workflows, and HTTP- and gRPC-based communication. We also study nine scenarios spanning three representative temporal fault patterns.

\subsubsection{Construct Threats to Validity}
Construct threats concern whether our metrics capture temporal-triggering effectiveness. We mitigate this threat by reporting \emph{CW}, \emph{PS}, \emph{Prem}, \emph{Miss}, and \emph{Mult} to separate failure modes and by applying benchmark-specific validation rules for post-effect scenarios across repeated trials. These metrics assess injection precision but do not directly quantify user or business impact.

\section{Related Work}
Chaos engineering commonly relies on fault injection to expose latent resilience bugs in distributed systems and cloud-native microservices~\cite{ChaosEngineeringBasic, Netflix-ICSE, ChaosMonkey, ChaosMesh, ChaosBlade}. Previous work has advanced FI through scenario selection and failure analysis~\cite{Coverage-ICC, FailureModeAnalysis, CausalFaultInject, RiskAnalysis}, reproducing realistic distributed failures~\cite{SyscallChaos, Rainmaker, FeedbackDrivenFailureReproduction, SlowFault}, and microservice-oriented testing based on lineage-driven search, fitness-guided prioritization, request-level injection, recovery testing, and resilience profiling~\cite{ldfi, LDFI-Netflix, IntelliFI, MicroFI, Filibuster, MicroRes, FastFI}. However, these approaches primarily determine where and what faults to inject, rather than when faults should become active during a distributed request execution.

Proxy-based FI reduces application-level instrumentation and supports resilience testing in microservice deployments, as demonstrated by Gremlin~\cite{Gremlin}. To the best of our knowledge, 3MileBeach~\cite{3milebeach} is the only prior work that explicitly models temporal prerequisites for fault activation in microservice fault injection. It introduces temporal fault injection by guarding fault activation with temporal prerequisites over inter-service message flows, enabling timing-sensitive resilience bugs.
In contrast, SequenceFI targets practical deployment in cloud-native microservices through non-intrusive sidecar enforcement, lightweight temporal-evidence propagation, and occurrence-aware temporal fault injection. 

\section{Conclusion}
This paper presents SequenceFI, a non-intrusive framework for temporal fault injection in microservice systems. By observing message-level send/receive events, propagating compact temporal evidence, and synthesizing occurrence-sensitive \texttt{After} guards from traces, SequenceFI enables faults to be triggered at precise temporal windows without modifying application code or serialization libraries. Evaluations on four benchmarks demonstrate that SequenceFI makes temporal fault injection practical, precise, and lightweight for microservice resilience testing. Future work will extend SequenceFI to support richer temporal predicates and protocols.

% \section{Data Availability}
% Our code and benchmarks are publicly available at: https://anonymous.4open.science/r/SequenceFI-451E/.

\bibliographystyle{IEEEtran}
\bibliography{refs}

\end{document}